\documentclass{article}

\usepackage{PRIMEarxiv}

\usepackage{amsfonts}       
\usepackage{amsmath}
\usepackage{array}          
\usepackage{booktabs}       
\usepackage{caption}        
\usepackage{enumitem}
\usepackage{fancyhdr}       
\usepackage[T1]{fontenc}    
\usepackage{graphicx}       
\usepackage{hyperref}       
\usepackage[utf8]{inputenc} 
\usepackage{lipsum}
\usepackage{microtype}      
\usepackage{multirow}       
\usepackage{nicefrac}       
\usepackage{svg}
\usepackage{url}            
\usepackage{wrapfig}
\usepackage{tikz}
\usetikzlibrary{positioning}

\graphicspath{{media/}}     
\title{BIOPTIC B1: a target-agnostic potency-based small molecules search engine
}

\author{
  Vlad Vinogradov \\
  Optic Inc. \\
  \texttt{vladvin@optic.inc} \\ 
  \And
  Ivan Izmailov \\
  Optic Inc. \\
  \texttt{smthngslv@optic.inc}  \\
  \And
  Simon Steshin \\
  Optic Inc. \\
  \texttt{simon.steshin@optic.inc}  \\
  \And
  Kong T. Nguyen \\
  Optic Inc. \\
  \texttt{kong@optic.inc}  \\
}

\begin{document}
\maketitle

\begin{abstract}
Recent successes in virtual screening have been made possible by large models and extensive chemical libraries. However, combining these elements is challenging: the larger the model, the more expensive it is to run, making ultra-large libraries unfeasible. To address this, we developed a target-agnostic, potency-based molecule search model, which allows us to find structurally dissimilar molecules with similar biological activities. We used the best practices to design fast retrieval system, based on processor-optimized SIMD instructions, enabling us to screen the ultra-large 40B Enamine REAL library with 100\% recall rate. We extensively benchmarked our model and several state-of-the-art models for both speed performance and retrieval quality of novel molecules.
\end{abstract}


\section{Introduction}
\label{sec:intro}
Drug discovery and development process consists of hit identification, hit to lead, lead optimization, and preclinical development stages. During this process, scientists search for ligands that interact with the targets of interest, and exhibit desired biological activity with favorable pharmacokinetics and safety profiles. Despite technology improvements and huge investments in the pharmaceutical industry, the current success rate of clinical drug candidates remains quite low with only 10\% of drug candidates become marketed drugs\cite{mullard2016parsing}. One of the main reasons of drug candidate failures is the lack of efficacy\cite{sun202290}. How can we significantly improve this process?

One of the modern approaches to drug discovery is virtual screening. This procedure runs a statistical algorithm on a large number of molecules, acting as a filter to narrow down the list of potential hits for experimental testing\cite{muratov2020qsar}. There have been several recent successes in virtual screening\cite{atomwise2024ai, wong2024discovery}, thanks to the large amounts of data that have become available in the last decades, which are required by data-greedy modern algorithms. Another key advantage virtual screening is its ability to scan huge libraries, which is not feasible in standard high-throughput screening (HTS) limited to hundreds of thousands of molecules. Advancement of synthetic chemistry methodologies and automation have led to the generation of huge virtual libraries, such as Enamine REAL Space\cite{grygorenko2020generating}, consisting of 40 billion different potential molecules, which is 5 orders of magnitude larger than the size of most HTS screening decks\cite{aldewachi2021high}.

Combining powerful data-greedy algorithms and ultra-large screening libraries poses a significant challenge. It is realistic to apply fast filters (like Lipinski's Rule of 5\cite{lipinski1997experimental}, Rule of CNS drugs\cite{rankovic2015cns}, etc) to huge libraries or to use neural networks to screen small datasets, but screening huge libraries using neural networks takes too much time and money. ML-based drug discovery is a blooming field with a large number of publications. Most of the publications focus on successful case studies rather than how to scale model predictions to huge libraries.

Stepping out of drug discovery, the modern world is all about data, and many industries have already transformed to handle large datasets and are reaping the benefits. In practice, modern recommendation systems (YouTube, Instagram, Twitter) and search engines (Google) are heavily bounded by computational resources, so they have invented new methods to make searches and predictions extremely fast and meaningful, leading to increased conversions and revenues\cite{covington2016deep, fbScalingInstagram, google}.

In this work, we apply the best practices of the most established software engineering tools --- recommendation systems and search engines --- to scale virtual screening of ultra-large libraries and demonstrate its performance in efficacy prediction on BindingDB and its speed on the Enamine REAL Space 40B library.


\section{Related works}
\subsection{Deep docking}
Docking is a traditional technique in computational chemistry that allows the simulation of ligand binding to a receptor under certain assumptions. Even with physically unrealistic assumptions, docking is not feasible for huge libraries\cite{gaillard2018evaluation}, so there are efforts to improve its performance\cite{gentile2020deep}.

In the recent papers\cite{gentile2021automated}, the authors suggest a hybrid architecture that uses docking to create a synthetic dataset and then trains a fast neural network on that dataset to make predictions for the entire library. This approach was fast enough to screen a 40 billion library.

In our work, we use only ground truth binding data, without relying on docking simulations. Furthermore, deep docking approach requires a known target structure and re-training the model for each target, which consumes a lot of resources. Instead, we employed search engine techniques to develop a global target-agnostic model that captures ligands’ properties, which are independent of targets. This allows us to have a single model that can find molecules with similar activities even when they are structurally dissimilar.

\subsection{DrugClip}
The recent paper DrugClip\cite{gao2024drugclip} represents the first attempt to create a retrieval system for binding affinity prediction. It employs a contrastive CLIP-like architecture to train a model on protein-ligand pairs, uses the UniMol 3D encoder, and suggests a new type of augmentation.

While it is a great approach to handle ultra-large libraries, DrugClip requires structural knowledge of proteins, which is not always available. Instead, we aimed to create a target-agnostic system capable of finding molecules with similar activity for all possible targets simultaneously.

\subsection{Chemprop}
Chemprop\cite{yang2019analyzing, heid2023chemprop} is a graph neural network that uses the Directed Message Passing Neural Network (D-MPNN) architecture. What makes this work outstanding is its extensive practical validation: the model has been successfully used to discover various antibiotics and senolytics\cite{stokes2020deep, wong2024discovery, wong2023discovering}. Since real-life success is the ultimate benchmark for any predictive model, Chemprop stands out as one of the few models that have been tested and proven in practical applications.

However, Chemprop is a supervised learning `model trained to predict specific properties rather than to search for molecules with similar activities, and it was not developed for integration with huge libraries. Nevertheless, it remains a solid competitor in terms of accuracy. Therefore, we selected it as a benchmark for comparison in our study.

\section{Our contribution}
\begin{itemize}
    \item We developed a single global model for target-agnostic potency-based molecule search, which allows finding structurally dissimilar molecules with similar activities.
    \item We extensively benchmark our model against several state-of-the-art (SOTA) models in hit identification settings for their ability to find novel, structurally dissimilar molecules with similar activities.
    \item We developed an efficient system based on processor-optimized SIMD instructions and deployed it on a cluster of machines connected to a matrix of SSD-disks that make it feasible to scan huge 40B libraries super fast.
\end{itemize}

\section{Method}
In this section, we describe the model's architecture and the search engine design. The benchmarks and demonstrations of the system are presented in the next section.

\subsection{Modeling}
\begin{figure}[h]
    \centering
    \includegraphics[width=0.8\textwidth]{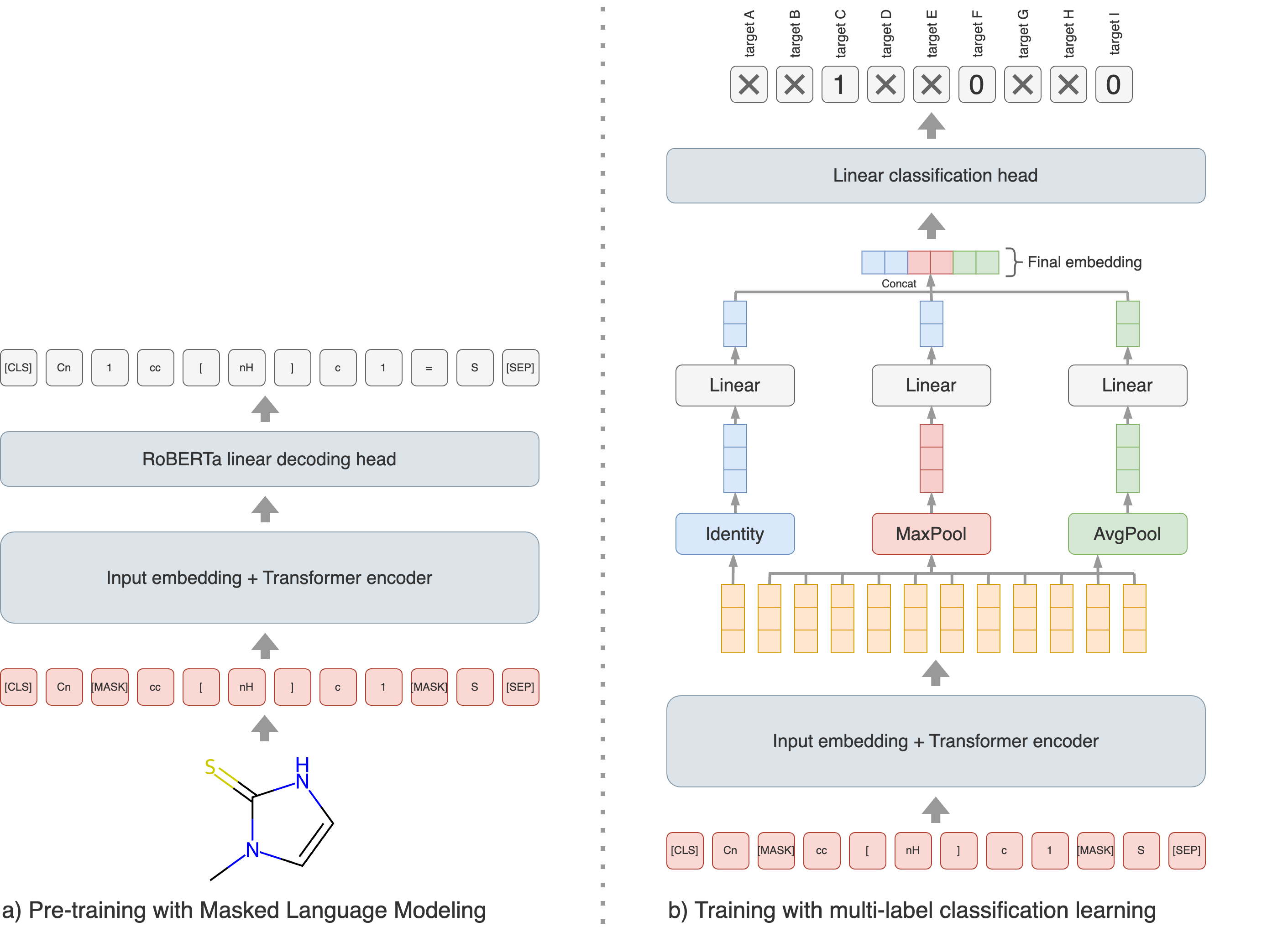}
    \caption{
        BIOPTIC’s training method includes two phases. In the initial phase (a),  a transformer neural network is pre-trained in a RoBERTa-style manner using masked language modeling on SMILES strings of molecules. In the next phase (b), the transformer model, augmented with a multi-pooling module and linear projection layers, is fine-tuned to predict which targets a given molecule is active to, using a masked binary cross-entropy loss function. A label is 1 if the molecule is active against a certain target and 0 otherwise. We mask out the unlabeled drug-target pairs.
    }
    \label{fig:nn_arch}
\end{figure}

Our model is a SMILES-based transformer \cite{vaswani2017attention}, which we train in two phases. The Byte-Pair Encoding (BPE) tokenization \cite{sennrich2016neural} is applied to SMILES strings at each phase. During the first phase, the model is pre-trained in an unsupervised fashion on a large corpus of unlabeled molecular data from PubChem and Enamine REAL Space. The training is performed via a masked language modeling following the procedure from RoBERTa \cite{liu2019roberta} (see Figure \ref{fig:nn_arch}a). This phase allows the model to learn the SMILES grammar.

In the second stage, the transformer model is augmented with a multi-pooling module that aggregates hidden layer output in three different ways: taking classification token output as is, and taking maximum and average along the sequence axis of all the remaining tokens. These aggregated vectors are passed into their corresponding linear layers for projection to a lower dimension following up with concatenation to form the final embedding. This embedding is further normalized and passed to the classification layer where the task is to predict which targets a given molecule is active to. Here, the model is trained on BindingDB. As the majority of labels are not available for the BindingDB drug-target pairs, we mask them out to not introduce additional noise into the training (see Figure \ref{fig:nn_arch}b).

We use binary cross-entropy loss with masking as an objective for learning potency-based molecular representations:
\begin{equation}
    L = \sum _{i=1}^{B}{\sum _{j=1}^{K}{m_{ij} \left[y_{ij} \cdot \log \hat{y}_{ij} + (1-y_{ij}) \cdot \log(1-\hat{y}_{ij}) \right]}}
\end{equation}
for a batch of size \(B\), where \(\hat{y}_{ij}\) is the prediction of the corresponding molecule \(i\) being active to target \(j\) (calculated from the logits of the classification layer after applying sigmoid function), \(y_{ij}\) is the corresponding ground truth value (0 or 1), and \(K\) is the total number of targets used in training. \(m_{ij}\) represents a masking indicator, which doesn't penalize the model on unlabeled data pairs, allowing it to learn from labeled drug-target interactions only. 

\subsection{Optimization}
The chosen model is RoBERTa, with a vocabulary size of 500, 6 hidden layers with a hidden size of 384, 8 attention heads, and an intermediate size of 1024, while the rest of the parameters are as described in \cite{liu2019roberta}. The model has a total of 8.7 million parameters. Initially, it is pre-trained in an unsupervised fashion on a large corpus of data comprising 115 million unique molecules from PubChem and a random set of 48 million molecules from the Enamine REAL Space library, totaling over 160 million molecules. The training is performed using masked language modeling following the procedure from RoBERTa \cite{liu2019roberta} (Figure \ref{fig:nn_arch}).

After pre-training, the hidden states of the last layer are extracted from the CLS token, and the rest of the tokens are max and average pooled. Three independent linear layers with an output size of 20 are then applied. The reduced representations are concatenated into a vector of size 60. The embedding dimensionality is set to 60 as a trade-off between quality and storage requirements. Embeddings are L2-normalized and passed to the classification linear layer, where the number of classes depends on the target-specific data split and averages around 6700 classes. For demonstrating performance with Hi splits, BIOPTIC is optimized on each data split for a binary classification objective for 300 epochs with the LARS optimizer \cite{you2017large}, using parameters: \(lr=0.1\), weight decay of \(1e-3\), \(\epsilon=1e-8\) and a trust coefficient of \(1e-3\). To enlarge our dataset, we utilize a randomization of the SMILES strings and random masking of 15\% of tokens in 30\% of cases during BIOPTIC training.

\subsection{Ultra-fast molecules retrieval system}
\begin{figure}[h]
    \centering
    \includegraphics[width=0.9\textwidth]{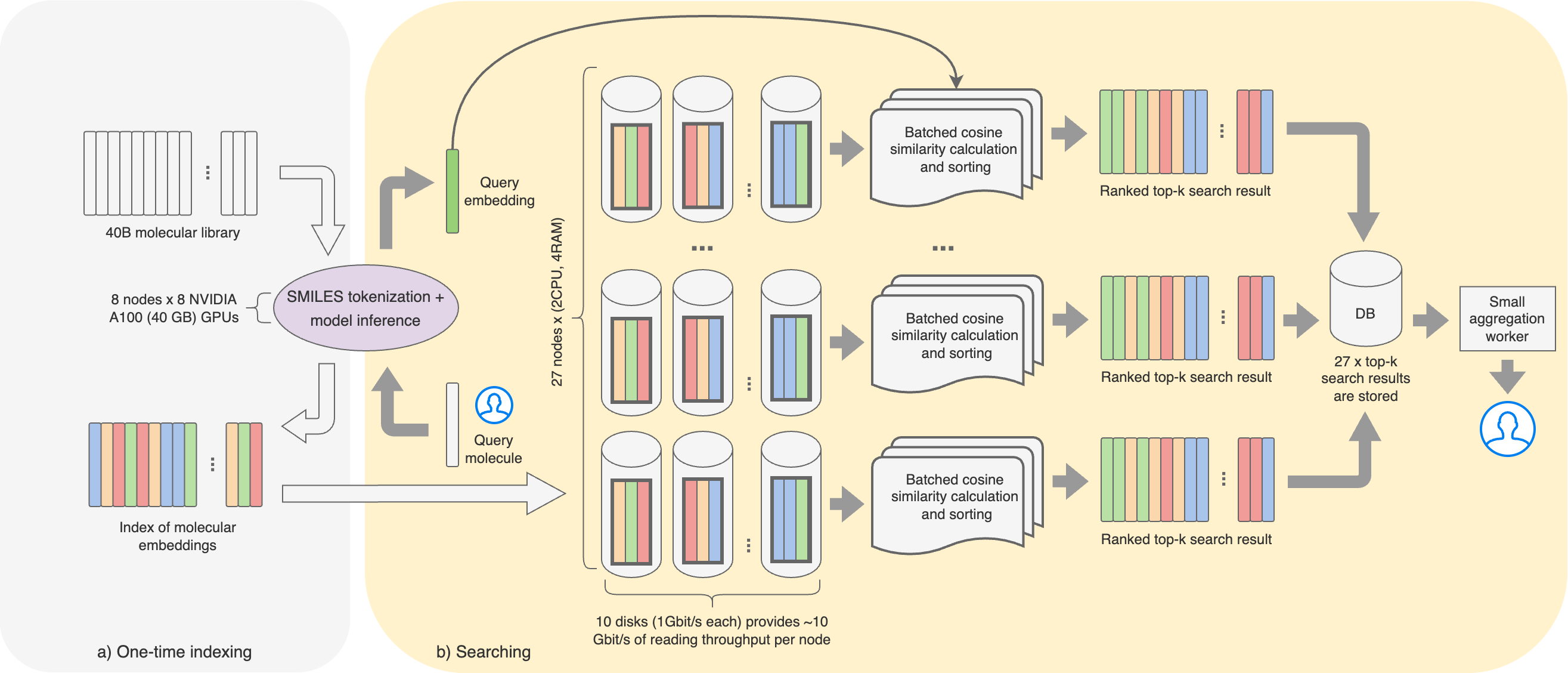}
    \caption{BIOPTIC molecular retrieval system design. The infrastructure is configured for processing the Enamine REAL Space 40 billion library. It includes a) a one-time indexing job performed on a cluster of NVIDIA GPU machines, and b) a searching job running on a cluster of CPU machines connected to a matrix of storage disks. The search operation allows a user to search multi-billion molecular libraries within seconds (see section \ref{sec:retr_perf}).}
    \label{fig:retrieval_system_design}
\end{figure}

The neural network produces molecular embeddings as \(float16\) value vectors in a 60-dimensional space. The number of dimensions is tuned to balance quality and retrieval speed. With these compact representations, we construct an index of a given molecular library and store it on SSD disks. To perform a search, we calculate the cosine similarity of a query molecule's embedding to each of the indexed molecules' embeddings in a brute-force fashion. Cosine similarity is implemented in C++ as matrix multiplication using Single-Instruction-Multiple-Data (SIMD) instructions. The main advantage of this method is the 100\% recall, meaning that if there exists an item in the index similar to the query, it must be found. This is crucial for large-scale libraries with billions of molecules.

For the Enamine REAL Space of 40 billion molecules, the inference of the neural network is performed on an AWS SageMaker cluster of 8 machines, each with 8 x NVIDIA A100 GPUs in \(float16\) precision. Processing also includes standardized SMILES tokenization using the Tokenizers library \cite{huggingface_tokenizers}. The full search job comprises the prediction steps and top-k selection steps running on multiple machines in parallel, followed by a top-k aggregation step performed using a PostgreSQL database on a single machine (see Figure \ref{fig:retrieval_system_design} for the full architecture overview). Multi-node parallel processing is an essential component of the design because reading molecular embeddings from SSD disks is a bottleneck. However, we opt out of using RAM for this task, as it would be much more expensive when keeping almost 5 TB of embeddings for Enamine REAL Space 40B in RAM. The search infrastructure is set up as follows: 27 nodes with 2-core CPUs and 4 GB RAM connected to 270 SSD disks of 21 GB each, with 1 Gbit/s of throughput per disk. With this infrastructure, we are able to cost-efficiently host clients willing to screen multi-billion molecular libraries within seconds (see section \ref{sec:retr_perf} for speed evaluation).

\section{Results}
\subsection{Hi-benchmark}

The ultimate goal of model development is not to excel at synthetic standard benchmarks retrospectively, but to provide meaningful chemical starting points for prospective drug discovery projects. We found most benchmarks unsuitable for this task, as they test models on a chemical space very similar to the training set, which was shown to the models during the training phase\cite{steshin2024hi}. This is contrary to hit identification stage, as we aim to apply the model in novel chemical spaces, and we do not find trivial modifications of existing drugs exciting.

For these reasons, we used Hi-splitter\cite{steshin2024hi} (Hi as in Hit Identification) and extended the Hi benchmark. This algorithm splits the dataset into training and holdout sets in such a way that no molecule in the test set has an ECFP4 Tanimoto similarity > 0.40 to any training molecule, representing novel chemical space (see Appendix \ref{appendix_hisplit_demo} for test set nearest neighbors). To perform well in these settings, models must generalize, as simple memorization would be insufficient for good results.

We focus on binding affinity, which is not the focus of the Hi benchmark, so we created our own benchmark. We selected seven very different targets representing four protein families (see Appendix \ref{appendix_datasets_analysis}). For each target, we created a separate dataset with its own training and holdout set. This approach helped us evaluate models extensively because we aimed to develop a single model that finds structurally dissimilar molecules with the same activities across various targets.

\subsection{Data}
We discuss data preprocessing in the Appendix \ref{appendix_data_preprocessing}, but briefly: we took BindingDB, preprocessed it, and binarized the binding affinity values such that all Ki, Kd, IC50 less than 10 $\mu$M are considered positive. After that, we created seven benchmarks from it. For each benchmark, we ran Hi-splitter to split the dataset into training and holdout sets and further split the holdout set into test and validation sets in a random fashion. To account for differences in class balance for different targets (there are many more actives than inactives for some targets, but not for others), we downsampled the test and validation datasets to achieve 1:1 activity balance. We also ensured that no molecule in the training set had an ECFP4 Tanimoto similarity > 0.40 to the test and validation sets.

\subsection{Hi-benchmark results}
\label{sec:hi_bench_results}
We benchmarked classic and SOTA models for their ability to identify dissimilar molecules with similar activity. For baselines, we selected three models: Chemprop, gradient boosting on ECFP4 fingerprints, and plain ECFP4 2048-bit fingerprint similarity search. The first two are popular choices for molecular property prediction, but they are single-target models and require retraining for each new target. The last is the standard method for searching for similar molecules.

To ensure a fair comparison, we conducted a thorough hyperparameter search using a validation set for each model (see Appendix \ref{appendix_hyperopt}) and calculated metrics on a test set that is dissimilar to the training set. The results are in the table \ref{tab:benchmark}.

For BIOPTIC and fingerprint search, we used 10 most active compounds from the training set as queries based on the analysis performed in section \ref{sec:queries-effect}. To get QSAR predictions from the model, we calculated cosine similarity between embeddings (or fingerprints) of queries and test set molecules. We ranked test molecules by maximal similarity to the queries, with the most similar at the top of the ranking.

\begin{table}[ht]
\centering
\begin{tabular}{lcccccc}
\toprule
\textbf{Target} & \textbf{Model} & \textbf{ROC AUC} & \textbf{AP} & \textbf{Precision@100} & \textbf{R-Precision} \\
\midrule
\multirow{5}{*}{ACHE} & Chemprop & \textbf{73.2} & \textbf{76.8} & \textbf{91.0} & \textbf{67.5} \\
                       & GB & 67.1 & 65.2 & 75.0 & 61.8 \\
                       & FP & 61.4 & 57.2 & 59.0 & 60.2 \\
                       & BIOPTIC (last) & 66.2 & 66.0 & 77.0 & 63.3 \\
                       & BIOPTIC (best) & 67.2 & 68.1 & 80.0 & 65.2 \\
\midrule
\multirow{5}{*}{AA2AR}  & Chemprop & 68.5 & 68.7 & 67.0 & 64.4 \\
                       & GB & 57.5 & 54.1 & 59.0 & 55.8 \\
                       & FP & 53.7 & 51.9 & 52.0 & 53.3 \\
                       & BIOPTIC (last) & 72.9 & \textbf{73.5} & \textbf{79.0} & 65.5 \\
                       & BIOPTIC (best) & \textbf{74.4} & 71.7 & 76.9 & \textbf{65.9} \\
\midrule
\multirow{5}{*}{DRD2}  & Chemprop & 59.6 & 55.4 & 53.0 & 54.9 \\
                       & GB & 64.2 & \textbf{64.8} & 65.0 & 63.1 \\
                       & FP & 49.5 & 53.5 & 53.5 & 47.2 \\
                       & BIOPTIC (last) & 65.9 & 60.3 & 63.0 & 62.6 \\
                       & BIOPTIC (best) & \textbf{69.1} & 63.3 & \textbf{68.9} & \textbf{65.9} \\
\midrule
\multirow{5}{*}{EGFR}  & Chemprop & 82.0 & \textbf{85.2} & 94.0 & \textbf{77.3} \\
                       & GB & 73.3 & 77.7 & 92.0 & 67.8 \\
                       & FP & 40.9 & 45.0 & 39.0 & 41.7 \\
                       & BIOPTIC (last) & \textbf{82.5} & 84.0 & 95.0 & 75.4 \\
                       & BIOPTIC (best) & 82.4 & 84.1 & \textbf{95.9} & 75.4 \\
\midrule
\multirow{5}{*}{5-HTT}   & Chemprop & \textbf{83.4} & \textbf{82.3} & \textbf{74.0} & \textbf{74.5} \\
                       & GB & 47.5 & 48.9 & 49.0 & 49.0 \\
                       & FP & 49.8 & 48.9 & 47.0 & 47.1 \\
                       & BIOPTIC (last) & 78.3 & 76.0 & 72.0 & 71.6 \\
                       & BIOPTIC (best) & 80.0 & 78.5 & 72.0 & 71.5 \\
\midrule
\multirow{5}{*}{JAK2}  & Chemprop & \textbf{81.1} & \textbf{83.0} &\textbf{ 70.0} & \textbf{77.3} \\
                       & GB & 56.5 & 62.1 & 52.0 & 54.7 \\
                       & FP & 51.4 & 49.1 & 51.0 & 47.6 \\
                       & BIOPTIC (last) & 77.8 & 78.9 & 68.0 & 69.0 \\
                       & BIOPTIC (best) & 76.7 & 78.3 & 66.0 & 73.8 \\
\midrule
\multirow{5}{*}{KCNH2} & Chemprop & \textbf{73.3} & \textbf{71.8} & \textbf{86.0} & \textbf{65.6} \\
                       & GB & 64.0 & 65.7 & 83.0 & 60.7 \\
                       & FP & 54.2 & 53.6 & 53.0 & 54.0 \\
                       & BIOPTIC (last) & 61.4 & 59.2 & 61.0 & 58.2 \\
                       & BIOPTIC (best) & 63.6 & 61.5 & 68.0 & 59.7 \\
\bottomrule
\end{tabular}
\caption{Benchmark results of the models. Here, checkpoints for Chemprop and Gradient Boosting models are picked at the best Average Precision metric on validation while BIOPTIC was tested in two scenarios: last checkpoint at the end of the training and best checkpoint according to Average Precision on validation. The best result is in bold. See the metrics description in the Appendix \ref{appendix_metrics}.}
\label{tab:benchmark}
\end{table}

\subsection{Effect of queries selection}
\label{sec:queries-effect}
\begin{figure}
  \centering
  \includegraphics[width=0.6\textwidth]{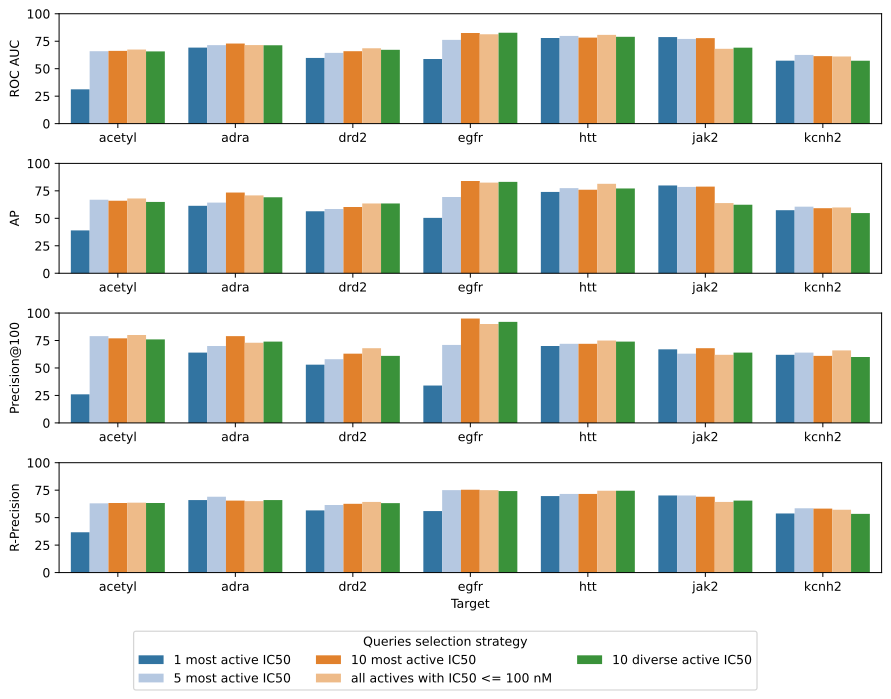}
  \caption{Impact of query selection strategy on model performance.}
  \label{fig:queries-effect}
\end{figure}

In order to investigate the effect of query selection, we evaluated the BIOPTIC model at the last training checkpoint for each target with different query selection strategies that include:
\begin{itemize}
    \item 1 most active (according to IC50 value)
    \item 5 most active (according to IC50 value)
    \item 10 most active (according to IC50 value)
    \item all actives with IC50 \(\leq\) 100 nM
    \item 10 diverse active molecules (based on agglomerative clustering into 10 clusters)
\end{itemize}

The results are in Figure \ref{fig:queries-effect}. All queries are selected from the training set. The "1 most active" strategy performs much worse on ACHE and EGFR, potentially due to noise in the data. The "5 most active," "10 most active," and "10 diverse active molecules" strategies are generally good.

\subsection{Retrieval performance on ultra-large molecular libraries}
\label{sec:retr_perf}
We designed our system with the need to screen ultra-large libraries in mind. To test the system, we evaluated speed performance under real-life conditions for screening 40 billion Enamine REAL compounds. We measured three scenarios:
\begin{enumerate}
    \item \textbf{Scan}: a one-time embedding extraction for all the molecules, performed with GPUs.
    \item \textbf{Search}: given query molecules and extracted embeddings, find the top-k molecules with similar activity in the library, performed with CPUs only.
    \item \textbf{RAM Search}: the same as Search, but embeddings are stored in RAM instead of slower SSDs. Instead of returning the best top-k, the search returns scores for the whole library.
\end{enumerate}
The scan operation runs at 30,000 molecules per second on a single NVIDIA A100 GPU and can be straightforwardly parallelized using more GPUs. For conducting search operations on 40 billion molecules, we deployed a distributed system of 27 cheap nodes with 2-core CPUs and 4 GB RAM (AWS’s c7g.large instances) connected to 270 SSD disks of 21 GB each, with 1 Gbit/s throughput per disk.

\begin{figure}[h]
    \centering
    \includegraphics[width=0.5\textwidth]{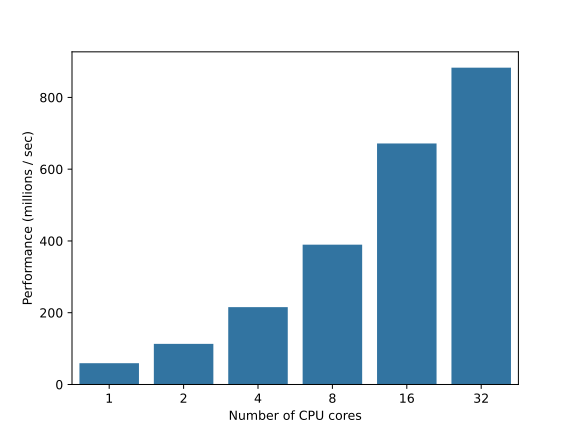}
    \caption{
        RAM search performance grows linearly with the number of cores.
    }
    \label{fig:simd_scaling}
\end{figure}

We found this step to be memory-bound by SSD throughput, so we measured performance for the in-RAM scenario. If all the embeddings are already loaded in RAM (which might be the case to handle a lot of search queries), we can process up to 1 billion molecules per second on a single 32-core AWS Graviton3 processor, with the speed increasing with the number of cores according to the demonstration in (see Figure \ref{fig:simd_scaling}).

See all time measurements in Table \ref{tab:speed}.

\begin{table}[ht]
    \centering
    \begin{tabular}{c|c|c|c}
        & 1M library & Enamine REAL 6B & Enamine REAL Space 40B \\
        \hline
        Scan & \(500\) msec & \(51\) min & \(5.6\) hours \\
        Search & \(4\) msec & \(20\) sec & \(2\) min \(15\) sec \\
        RAM Search & \(1\) msec & \(6.6\) sec & \(43.9\) sec \\
    \end{tabular}
    \caption{Speed performance of the BIOPTIC search system. Each search is performed with one query.}
    \label{tab:speed}
\end{table}

\subsection{Scaffold hopping from the box}
\begin{figure}[h]
    \centering
    \includegraphics[width=0.8\textwidth]{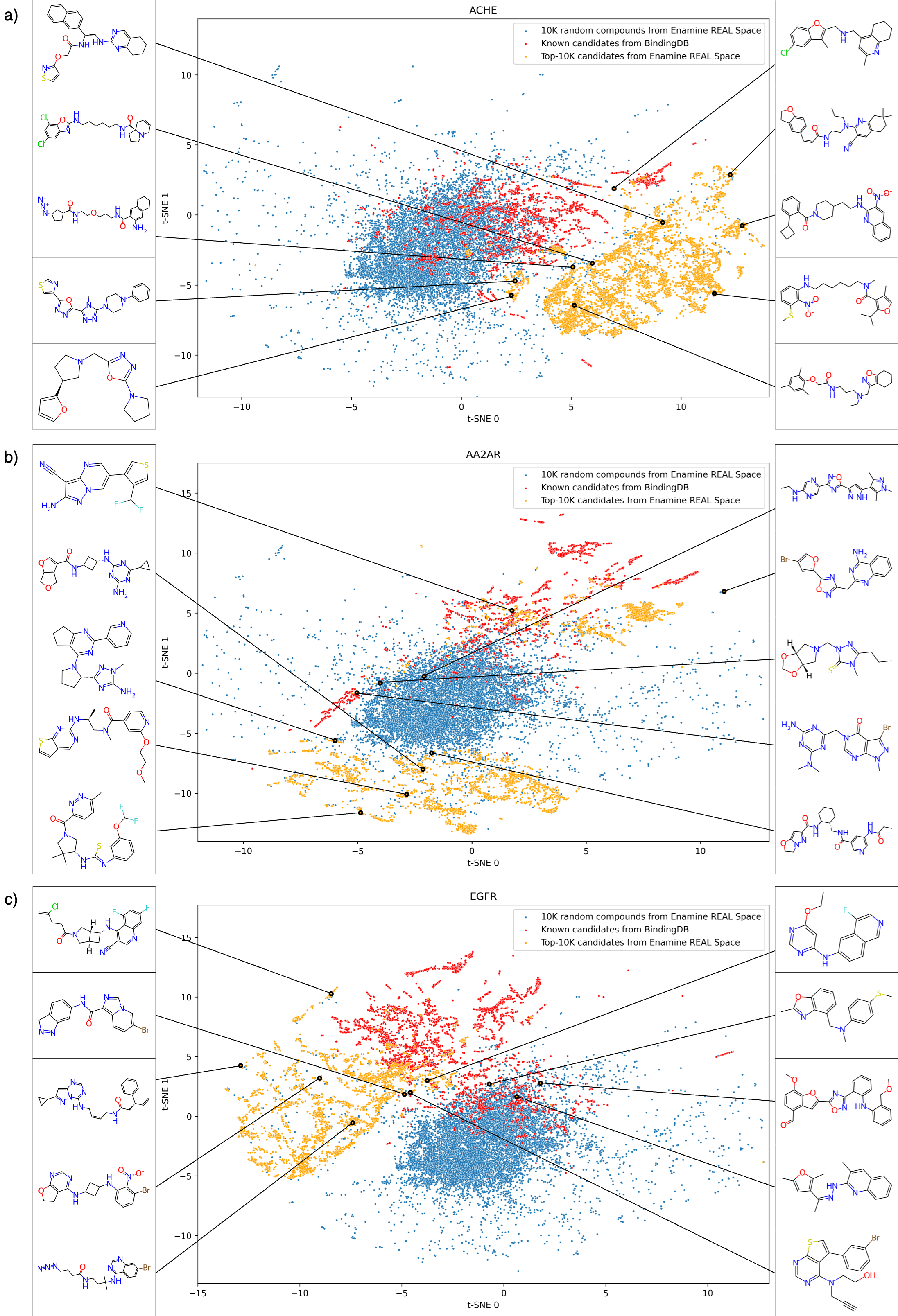}
    \caption{
        Diversity of top candidates retrieved with BIOPTIC from Enamine REAL Space 40 billion library for a) ACHE, b) AA2AR and c) EGFR.
    }
    \label{fig:retr_examples}
\end{figure}

To demonstrate diverse candidates selection performed with BIOPTIC, we chose ACHE, AA2AR and EGFR as targets and ran search in Enamine REAL Space 40B library. We used 10 diverse active compounds as queries and retrieved top-10k candidates (selection lists). We also took 10k random compounds from Enamine REAL Space and all known molecules tested for binding to the corresponding targets from BindingDB. We then mapped Morgan fingerprints (ECFP4) of all the molecules into 2D coordinates with t-SNE. From each selection list, we then removed all molecules that are closer than 0.4 Tanimoto similarity to any BindingDB molecule and picked 10 diverse molecules with RDKit's MaxMinPicker. The results are shown in \ref{fig:retr_examples}. It's seen from the figures that BIOPTIC 1) prioritizes molecules from a widely span corner of the Enamine chemical space, 2) picks molecules structurally different from what is already known, 3) selects candidates of different scaffolds.

\section{Conclusion}
With the exponential growth in size of small-molecule screening libraries, it’s necessary to consider the efficiency of virtual screening models. A 40 billion compound library is already prohibitively large for many models, which prompted us to develop a ultra-fast retrieval molecular search engine that can find dissimilar molecules with similar activity. We designed and deployed our BIOPTIC system using best practices of search engines. As a result of that, our system can search a pre-indexed library of 40 billion compounds in 40 seconds using CPU-only machines.

We understand that in many cases, chemists are interested in finding structurally novel molecules instead of minor modifications of known binders. That’s why we benchmarked our system against several state-of-the-art models on their ability to find active molecules with ECFP4 Tanimoto similarity < 0.4 to the training set for a wide variety of proteins representing different families. We believe that our system can search for novel molecules with desirable biological activity in ultra-large libraries in real-time.

\bibliographystyle{unsrt}  
\bibliography{references}  

\newpage
\appendix
\section{Data preprocessing}
\label{appendix_data_preprocessing}
We constructed benchmarks using the database \texttt{BindingDB\_All\_202310}. This section discusses the preprocessing steps.

We removed approximately 40 SMILES entries containing the asterisk (*). After that, we binarized the activity values for Ki (nM), Kd (nM), and IC50 (nM). We began by dividing all the IC50 values by a coefficient of 2.3 following \cite{kalliokoski2013comparability}. Both Ki and IC50 values contain important information useful for training, but they are not directly comparable. We used the 2.3 coefficient, as it is the common difference between IC50 and Ki values in databases; we hypothesized this could improve training. Some values contained greater (">") and lesser ("<") signs, so we considered a molecule active if its activity was less than 10,000 nM, and inactive if its activity was greater than 10,000 nM. All ambiguous values (such as > 5,000 nM) were discarded. If a SMILES-Target pair had several activity values, we preferred one of the binary values in this order: Ki, Kd, IC50.

As targets, we selected fields \texttt{UniProt (SwissProt) Primary ID of Target Chain} and \texttt{UniProt (TrEMBL) Primary ID of Target Chain}. We removed all data points that had either both or none of these fields. After that, we standardized the SMILES with ChEMBL's molecular standardization pipeline \cite{bento2020open} and removed data points with the same SMILES and target, but conflicting binarized activity values. Finally, we eliminated PAINS alerts using the medchem library.

\section{Hi-split nearest neighbors}
\label{appendix_hisplit_demo}
\begin{figure}[htbp]
    \centering
    \begin{tikzpicture}
        \node[anchor=south west,inner sep=0] (image1) at (0,0) {\includegraphics[width=0.24\textwidth]{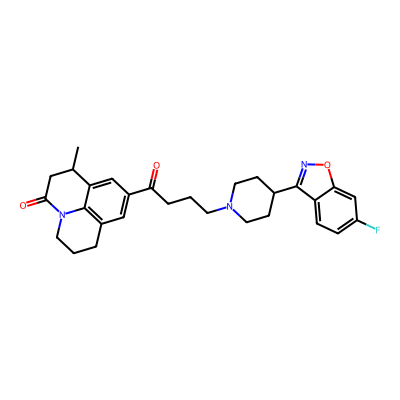}};
        \node[anchor=south west,inner sep=0] (image2) at ([xshift=0\textwidth]image1.south east) {\includegraphics[width=0.24\textwidth]{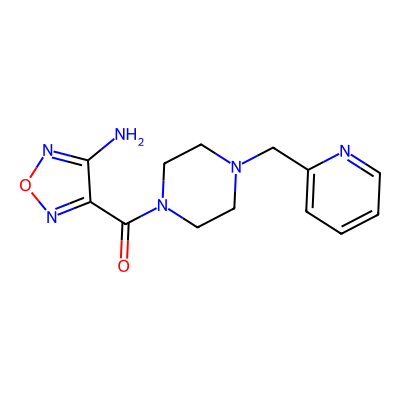}};

        \node[anchor=north west,inner sep=0] (image5) at ([yshift=-1cm]image1.north west|-image1.south) {\includegraphics[width=0.24\textwidth]{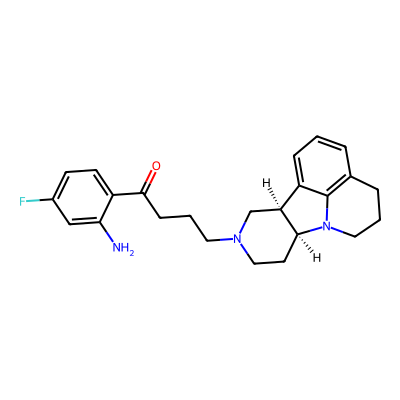}};
        \node[anchor=north west,inner sep=0] (image6) at ([yshift=-1cm]image2.north west|-image2.south) {\includegraphics[width=0.24\textwidth]{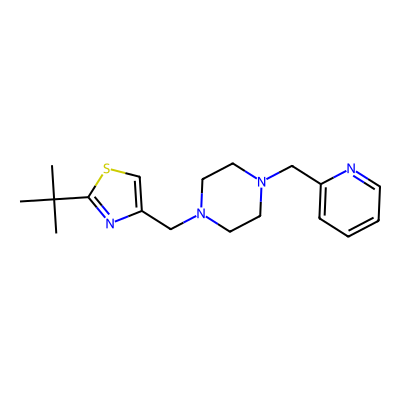}};

        \node[below=0.3cm of image1,font=\small] {Train molecule \textbf{1}};
        \node[below=0.3cm of image2,font=\small] {Train molecule \textbf{2}};
        \node[below=0.3cm of image5,font=\small] {Test neighbor of \textbf{1}};
        \node[below=0.3cm of image6,font=\small] {Test neighbor of \textbf{2}};
    \end{tikzpicture}
    \caption{The most similar test neighbors of the train molecules in the DRD2 dataset.}
    \label{fig:appendix_nearest_mols}
\end{figure}

\pagebreak
\section{Datasets analysis}
\label{appendix_datasets_analysis}
To create a dataset for a specific target, we selected all molecules with known activity to the target and split them into train and holdout sets using Hi-splitter. We further randomly split the holdout set into test and validation datasets. After that, we removed from the train set all molecules that have an ECFP4 Tanimoto similarity > 0.4 to those in the test and validation datasets. Next, we downsampled the test and validation datasets to achieve a 1:1 class balance. The detailed statistics are in the Table \ref{tab:targets}.

\begin{table}[!]
    \centering
    \begin{tabular}{|>{\centering\arraybackslash}p{3cm}|>{\centering\arraybackslash}p{1.6cm}|c|>{\centering\arraybackslash}p{3cm}|c|c|c|}
        \midrule
        \textbf{Name} & \textbf{Shorthand} & \textbf{UniProt ID} & \textbf{Family} & \textbf{Type} & \textbf{Actives} & \textbf{Inactives} \\
        \midrule
        Acetylcholinesterase & ACHE & P22303 & Carboxylesterase & Globular & 3396 & 2535 \\
        \midrule
        Adenosine A2A receptor & AA2AR & P29274 & G protein-coupled receptor (GPCR) & Membrane & 6165 & 1002 \\
        \midrule
        D2 dopamine receptor & DRD2 & P14416 & G protein-coupled receptor (GPCR) & Membrane & 7912 & 1243 \\
        \midrule
        Epidermal growth factor receptor & EGFR & P00533 & Receptor tyrosine kinase & Membrane & 8130 & 2715 \\
        \midrule
        Serotonin transporter & 5-HTT & P31645 & Solute carrier family & Membrane & 4452 & 783 \\
        \midrule
        Tyrosine-protein kinase JAK2 & JAK2 & O60674 & Non-receptor tyrosine kinase & Globular & 11986 & 1479 \\
        \midrule
        Potassium voltage-gated channel subfamily H member 2 & KCNH2 & Q12809 & Voltage-gated potassium channel & Membrane & 3644 & 5560 \\
        \midrule
    \end{tabular}
    \caption{Targets from BindingDB selected for evaluation. The selection strategy is based on including different families and types of proteins while maintaining enough active and inactive compounds. This table summarizes the total number of active and inactive compounds present in the BindingDB after all the filtering steps.}
    \label{tab:targets}
\end{table}

\section{Metrics description}
\label{appendix_metrics}
Virtual screening is used as a filter, to select the top rated molecules. The primary goal is to ensure that active molecules are ranked at the top, optimizing the screening efficiency. Here, we discuss several key metrics used for this purpose:

\subsection{ROC AUC}
The Receiver Operating Characteristic (ROC) Area Under the Curve (AUC) is a standard ranking metric. It represents the area under the curve plotting the True Positive Rate (TPR) against the False Positive Rate (FPR). An AUC of 0.5 indicates random ranking, while an AUC of 1.0 signifies perfect ranking. We used the \texttt{sklearn.metrics.roc\_auc\_score} to compute this metric.

\subsection{AP}
Average Precision (AP) score summarizes a precision-recall curve by calculating the weighted average of precision values at each threshold. The weights are determined by the increase in recall from the previous threshold. This metric provides a single-figure measure of quality across different recall levels and prefer models with early recognition. We used the \texttt{sklearn.metrics.average\_precision\_score} to compute this metric.

\subsection{Precision@100}
Precision@100 measures the fraction of active molecules within the top 100 molecules in the ranked list. This metric is particularly useful for assessing the effectiveness of virtual screening when focusing on the top candidates.

\subsection{R-Precision}
R-Precision is equivalent to Precision@R, where R is the number of total active molecules in the test set. This metric assesses the proportion of active molecules retrieved when considering a number of top-ranked molecules equal to the total number of actives. It helps in understanding the retrieval performance relative to the available actives.

\section{Hyperparameter tuning}
\label{appendix_hyperopt}
To ensure a fair comparison, we conducted hyperparameter tuning.

\subsection{Chemprop}
We used Chemprop 2.0.0 and performed 80 iterations of random search, sampling these parameters:
\begin{verbatim}
PARAMS_TO_SAMPLE = {
    "max_epochs": [60],
    "mp": ["bond", "atom"],
    "d_h": [150, 300, 600, 900, 1200, 2400],
    "depth": [3, 4, 5, 6],
    "dropout": [0.0, 0.0, 0.2, 0.3, 0.4, 0.5],
    "batch_norm": [False, True],
    "agg": ["mean", "sum", "norm"],
    "bias": [True, False],
    "init_lr": [1e-4],
    "max_lr": [1e-3, 3e-4, 1e-4],
    "final_lr": [1e-4],
    "batch_size": [32, 64, 96],
    "ffn_hidden_dim": [300, 600, 900],
    "ffn_n_layers": [1, 2, 3],
    "ffn_dropout": [0.0, 0.0, 0.1, 0.2, 0.3],
}
\end{verbatim}

The best checkpoint was selected based on the best validation average precision.

\subsection{GB}
We used GradientBoostingClassifier from scikit-learn 1.2.2. We trained it on ECFP4 2048 fingerprints. For each benchmark, we ran 30 iterations of random search, sampling these parameters:
\begin{verbatim}
params = {
    'n_estimators': [10, 50, 100, 150, 200, 250, 500],
    'learning_rate': [0.01, 0.1, 0.3, 0.5, 0.7, 1.0],
    'subsample': [0.4, 0.7, 0.9, 1.0],
    'min_samples_split': [2, 3, 5, 7],
    'min_samples_leaf': [1, 3, 5],
    'max_depth': [2, 3, 4],
    'max_features': [None, 'sqrt']
}
\end{verbatim}

The best checkpoint was selected based on the best validation average precision.
\end{document}